%% file: DISCO_-_Distributed_SDN_Controllers__under_submission_.tex
\documentclass[conference]{IEEEtran}

  \usepackage[pdftex]{graphicx}
  \usepackage{subfigure}
 \usepackage{amsmath}

  \graphicspath{{./figures/}}
  \DeclareGraphicsExtensions{.pdf,.jpeg,.png,.pdf}
  
\begin{document}

\title{DISCO: Distributed Multi-domain SDN Controllers}

\author{\IEEEauthorblockN{K\'{e}vin Phemius, Mathieu Bouet and J\'{e}r\'{e}mie Leguay}
\IEEEauthorblockA{Thales Communications \& Security\\
4 avenue des Louvresses, 92230 Gennevilliers, France\\ 
\{kevin.phemius, mathieu.bouet, jeremie.leguay\}@thalesgroup.com}
}

\hyphenation{}

\maketitle

\begin{abstract}
Modern multi-domain networks now span over datacenter networks, enterprise networks, customer sites and mobile entities. Such networks are critical and, thus, must be resilient, scalable and easily extensible. The emergence of Software-Defined Networking (SDN) protocols, which enables to decouple the data plane from the control plane and dynamically program the network, opens up new ways to architect such networks. In this paper, we propose DISCO, an open and extensible \emph{DIstributed SDN COntrol plane} able to cope with the distributed and heterogeneous nature of modern overlay networks and wide area networks. DISCO controllers manage their own network domain and communicate with each others to provide end-to-end network services. This communication is based on a unique lightweight and highly manageable control channel used by \textit{agents} to self-adaptively share aggregated network-wide information. We implemented DISCO on top of the Floodlight OpenFlow controller and the AMQP protocol. We demonstrated how DISCO's control plane dynamically adapts to heterogeneous network topologies while being resilient enough to survive to disruptions and attacks and providing classic functionalities such as end-point migration and network-wide traffic engineering. The experimentation results we present are organized around three use cases: inter-domain topology disruption, end-to-end priority service request and virtual machine migration.
\end{abstract}

\input{introduction}

\input{related}

\input{archi}

\input{implem}
\input{evaluation}
\input{conclusion}

\bibliographystyle{IEEEtran}
\bibliography{references}

\end{document}

%% file: introduction.tex
\section{Introduction} 

\begin{figure}[!t]
\centering
\includegraphics[width=0.8\linewidth]{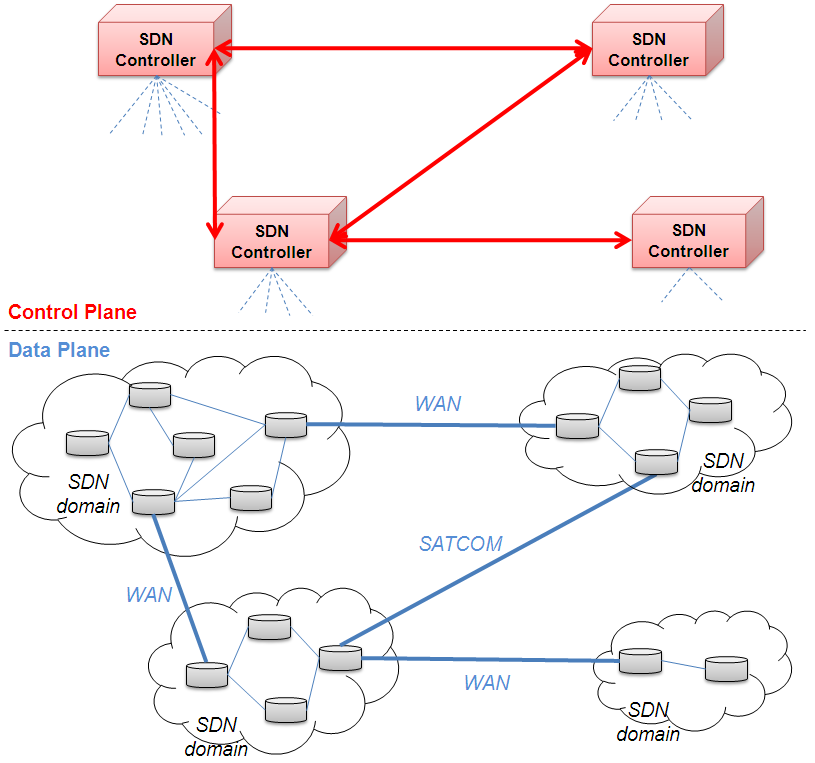}
\caption{Typical deployment of the multi-domain SDN control plane DISCO.}
\label{fig:deployment}
\end{figure}

Resilient, scalable and extensible networks are critical to interconnect datacenters, enterprise networks and even, potentially deployable, access networks. The Software Defined Networking (SDN) paradigm has emerged from the need to overcome the primary limitations of today's networks: complexity, lack of scalability and vendor dependence. It is based on three main principles: separation of software and physical layers, centralized control of information and network programmability. The ability to program networks, interact with network elements and manage a unified multi-vendor multi-technology environment enables service providers and network operators to innovate faster and to reduce operational and capital expenditures. SDN was first massively used in datacenters, network management being thus integrated to cloud platforms. It is now envisioned for multi-datacenter environments and multi-domain networks~\cite{b4}.\\
The SDN paradigm has emerged over the past few years through several initiatives and standards, FORCES~\cite{rfc3746} being one example. The leading SDN protocol in the industry is the OpenFlow protocol. It is specified by the Open Networking Foundation (ONF)~\cite{onf}, which regroups the major network service providers and network manufacturers. The majority of current SDN architectures, OpenFlow-based or vendor-specific, relies on a single or master/slave controllers, that is a physically centralized control plane. This centralization, adapted for datacenters, is not suitable for wide multi-technology multi-domain networks. In addition, the centralized SDN controller represents a Single Point Of Failure (SPOF), which makes SDN architectures highly vulnerable to disruptions and attacks~\cite{kreutz2013}.\\ 
Recently, proposals have been made to physically distribute the SDN control plane, either with a hierarchical organization~\cite{kandoo} or with a flat organization~\cite{hyperflow}. These approaches avoid having a SPOF and enable to scale up sharing load among several controllers. However, these distributed SDN control planes have been designed for datacenters, where controller instances share huge amount of information to ensure fine-grained network-wide consistency.\\
In this paper, we address multi-domain SDN networks, like the one presented in Fig.~\ref{fig:deployment}, which can be deployed to interconnect datacenters, enterprise networks, customer sites and mobile entities. They are generally decomposed into administrative or geographical domains interconnected with a large variety of network technologies from high-capacity leased lines to limited-bandwidth satellite links, or from costly but highly secured links to cheap but unsecured ones. The distributed and heterogeneous nature of these environments call for a distributed multi-domain network control plane which should be lightweight, adaptable to user or network requirements, and robust to failures. Current state of the art distributed SDN solutions are not suitable, as they do not provide a fine grain mean to control and adapt inter-controller information exchanges.\\
We propose DISCO, an open \emph{DIstributed SDN COntrol plane} for multi-domain SDN networks. It relies on a per domain organization, where each DISCO controller is in charge of an SDN domain, and provides a unique lightweight and highly manageable control channel used by \textit{agents} that can be dynamically plugged into the different domain controllers. The agents that we have developed share between the domains aggregated network-wide information and 
hence provide end-to-end network services. We demonstrate how DISCO's control plane dynamically adapts to heterogeneous network topologies while providing classic functionalities such as end-point migration and network-wide traffic engineering and being resilient enough to survive to disruptions and attacks. Contrary to state of the art distributed SDN control planes, DISCO well discriminates heterogeneous inter-domain links such as high-capacity MPLS tunnels and SATCOM interconnections with poor bandwidth and latency. We implemented DISCO on top of the Floodlight~\cite{floodlight} OpenFlow controller and the AMQP~\cite{amqp} protocol. To evaluate its performance, we show an evaluation of its functionalities on an emulated software defined network according to three use cases: inter-domain topology disruption, end-to-end priority service request and virtual machine migration.\\
The rest of this paper is organized as follows. First, Sec.~\ref{sec:related_work} analyzes related work. Then, Sec.~\ref{sec:disco} presents DISCO architecture composed of an intra-domain part and an inter-domain part. Our implementation of DISCO is explained in Sec.~\ref{sec:implementation}. The different use cases we considered and their evaluation are tackled in Sec.~\ref{sec:evaluation}. Finally, Sec.~\ref{sec:conclusion} concludes this paper.

%% file: related.tex
\section{Related work}
\label{sec:related_work}

Parallel open source initiatives such as NOX~\cite{nox}, Beacon~\cite{beacon}, Floodlight~\cite{floodlight}, Ryu~\cite{ryu} etc. Researchers mainly focused on improving the performance of a specific controller like Maestro~\cite{Cai2010}  and NOX~\cite{Gude2008} or demonstrating the improvement offered by OpenFlow against a classic L2 paradigm~\cite{Bianco2010}.\\
Several attempts have been done to tackle the problem of scaling SDNs. A first class of solutions, such as DIFANE~\cite{difane} and DevoFlow~\cite{devoflow}, address this problem by extending data plane mechanisms of switches with the objective of reducing the load towards the controller. DIFANE tries to partly offload forwarding decisions from the controller to special switches, called authority switches. Using this approach, network operators can reduce the load on the controller and the latencies of rule installation. DevoFlow, similarly, introduces new mechanisms in switches to dispatch far fewer `important' events to the control plane.\\
A second class of solutions proposes to distribute controllers. HyperFlow~\cite{hyperflow}, Onix~\cite{onix}, and Devolved controllers~\cite{devolved} try to distribute the control plane while maintaining a logically centralized using a distributed file system, a distributed hash table and a pre-computation of all possible combinations respectively. These approaches, despite their ability to distribute the SDN control plane, impose a strong requirement: a consistent network-wide view in all the controllers. On the contrary, Kandoo~\cite{kandoo} proposes a hierarchical distribution of the controllers based on two layers of controllers: (i) the bottom layer, a group of controllers with no interconnection, and no knowledge of the network-wide state, and (ii) the top layer, a logically centralized controller that maintains the network-wide state.\\
In addition,~\cite{tradeoff} analyzes the trade-off between centralized and distributed control states in SDN, while~\cite{placement} proposes a method to optimally place a single controller in an SDN network.\\
Recently, Google has presented their experience with B4~\cite{b4}, a global SDN deployment interconnecting their datacenters. In B4, each site hosts a set of master/slave controllers that are managed by a gateway. The different gateways communicate with a logically centralized Traffic Engineering (TE) service to decide on path computations. While BGP is used between border network elements to exchange routes, possibly with external service providers or operators, proprietary APIs and protocols are used between all the software pieces (gateways, controller, TE engine).\\
DISCO differs from state of the art solutions as it provides an open distributed control plane for multi-domain networks based on a unique message-oriented communication bus. Indeed, state of the art distributed control planes are not adaptable to heterogeneous network deployments. Most of them impose a consistent network-wide state in all controllers and thus generate large control traffic. On the contrary, DISCO separates intra-domain and inter-domain specific information. Furthermore, it currently features \textit{agents} for adaptive monitoring with regards to security restrictions and low bandwidth interconnections, and \textit{agents} for end-to-end QoS and mobility management.

%% file: archi.tex
\section{DISCO Architecture}
\label{sec:disco}
In this section, we present DISCO's controller architecture, detailing both intra and inter-domain modules. 

\subsection{Overall architecture}
\label{sec:architecture}

\begin{figure}[!t]
\centering
\includegraphics[width=0.95\linewidth]{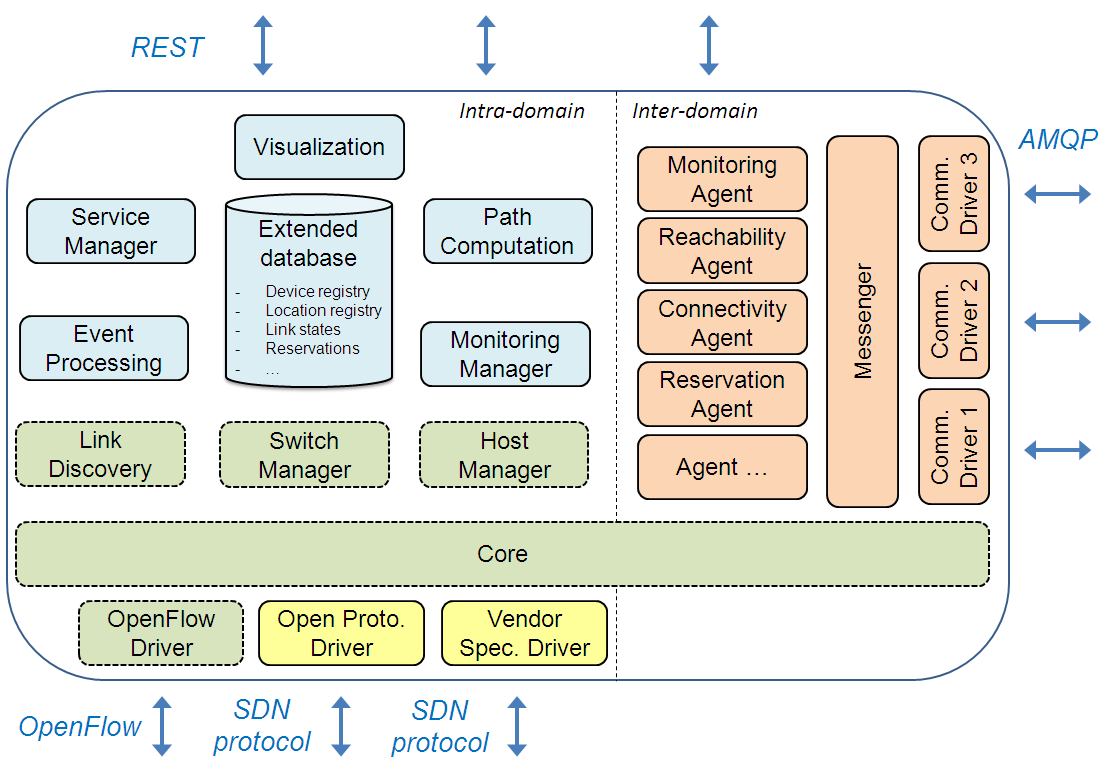}
\caption{DISCO Controller Architecture.}
\label{fig:architecture}
\end{figure}

DISCO is a distributed multi-domain SDN control plane which enables the delivery of end-to-end network services.
A DISCO controller is in charge of a network domain and communicates with neighbor domains to exchange 
aggregated network-wide information for end-to-end flow management purposes. \\
Fig.~\ref{fig:architecture} presents the architecture. It is composed of two parts: an intra-domain part, 
which gathers the main functionalities of the controller, and an inter-domain part, which manages the 
communication with other DISCO controllers (reservation, topology state modifications, disruptions, \ldots). 
In addition to this east-west interface, a controller has at least one southbound SDN interface used to 
push policies to the network elements and retrieve their status. Finally, a northbound interface enables 
to push management policies to the controller (e.g., service and user priorities), to manage SLAs and report 
network service status. 
A controller is composed of several modules managed by the \textit{Core} component. It enables to start, 
stop, update the modules and provides them with a communication bus. Our architecture leverage from 
classical off-the-shelf modules that SDN controllers provide~\cite{floodlight}
such as an \textit{OpenFlow driver} to implement the OpenFlow protocol, a \textit{switch manager} 
and \textit{host manager} to keep track of the different network elements, and \textit{link discovery} implementing
LLDP (Link Layer Discovery Protocol). In the rest of this section, we present the modules that we have 
specifically developed for intra-domain and inter-domain flow management.

\subsection{Intra-domain functionalities}

The intra-domain modules enable to monitor the network and manage flow prioritization so that the controller can compute the routes of priority flows based on the state of the different network parameters. The modules also enable to dynamically react to network issues (broken link, high latency, bandwidth cap exceeded, \ldots) by redirecting and/or stopping traffic according to the criticality of the flow. This work extends our previous work~\cite{bouet2012} demonstrated in a centralized architecture or intra-domain context.\\
The \emph{Extended Database} is a central component in which each controller stores all the intra-domain and inter-domain knowledge on network topology, monitoring and ongoing flows. All the modules and agents presented in the rest of this section either enrich or use this information, with the ultimate goals of taking actions on flows. This database and its associated model resemble to what the recent IETF group Interface to the Routing System (I2RS)~\cite{i2rs}
is trying to standardize.\\
The \emph{Monitor Manager} module gathers information such as the flow throughput on the switches using the appropriate messages described in the OpenFlow protocol specification (STATISTICS\_REQUEST and REPLY)~\cite{onf}. In addition, this module measures the one-way latency and packet loss rate on intra-domain links by sending at a given source node specially crafted Ethernet frames including a time-stamp, and retrieving them at a given destination node to measure the elapsed time. This method has been described in~\cite{phemius2013}. By doing these operations periodically, the controller maintains an up-to-date view of link and network devices performances in the \textit{Extended database}. For peering links with other domains, a simple Ping is used to estimate the round trip delay and packet losses.\\
The \emph{Events Processing} module keeps tracks of variations or saturation events. 
By setting up ceiling values, the controller can immediately react if 
a value goes out of bound. \emph{Events} can work with absolute values (e.g.,the total amount of dropped packets on a port) or relative values (e.g., the number of lost packets in the last second). \\
The \emph{Path Computation} module computes routes for flows from source to destination using a variation of the Dijkstra algorithm, that is with QoS metrics and taking into account existing reservations. If a link on the route is considered impaired (by consulting the \emph{Events Processor} module), a new route is computed and flow pre-emption mechanisms are applied if necessary. This module uses the \textit{Extended Database} to retrieve information about flow descriptions (e.g., priority, bandwidth and latency constraints), network topology (e.g., capacity, routing preferences) and monitoring
information. \\
The \textit{Service Manager} module is responsible for the management of network SLAs inside its domain. Upon reception of a service request and all along the network lifetime, it verifies, using other modules, the feasibility and respect of SLAs. It can receive requests from the northbound API or from neighboring domains for
end-to-end service provisioning.\\
On top of these modules, a GUI through the \emph{Visual Manager} module allows the 
visualization of the network and the interaction between the user and the modules. It gather information from many other modules of the controller to display relevant information to the network operator and provides parameterization capabilities (e.g., flow priorities, new events, new routes).  

\subsection{Inter-domain functionalities}
\label{sec:inter}

A DISCO controller communicates with neighbor domain controllers to exchange aggregated network-wide information.
They are composed of two key elements: (i) a \textit{Messenger} module which discovers neighboring controllers 
and maintain a distributed publish/subscribe communication channel, and (ii) different \textit{agents} that use this 
channel to exchange network-wide information with intra-domain modules. This way \emph{Path Computation} 
can learn for instance to which neighbor domain a packet has to be routed to reach a given host.

\subsubsection{Messenger}

This module implements a unique control channel between neighboring domains. It should support group and direct communications to exchange status information (link state, host presence) and request actions (e.g., reservations) from other controllers. The usual communication patterns used by IETF protocols such as Open Shortest Path First (OSPF) protocol, Resource Reservation Protocol (RSVP) and Border Gateway Protocol (BGP) should be supported, namely step-by-step diffusion (e.g.,distance vectors), network-wide flooding (e.g., link states), uncased queries (e.g., reservation requests), and publish/subscribe messages (e.g., route updates).\\
To meet these requirements, we have chosen the Advanced Message Queuing Protocol (AMQP)~\cite{amqp} as a base for 
the implementation of \textit{Messenger}. AMQP is an open standard and a thin application layer protocol for message-oriented middleware.
It offers built-in features for message orientation, queuing with priority, 
routing (including point-to-point and publish-and-subscribe), reliable delivery and security.
Due to the convergence of network and IT systems such as cloud management,
AMQP is an interesting solution being lightweight, highly controllable and software-oriented. 
It is, for instance, used in OpenStack~\cite{openstack} for loosely coupled communication between the different components.\\
AMQP is generally used in client-server mode. This means that \textit{Messenger} executes a server and uses clients to connect to the different servers of neighboring controllers. In this mode, \textit{Messenger} only help local agents
to exchange information with agents of neighbor domains, but does not provide communication support for network-wide exchanges. 
For this, it could relay the information for one domain to another by implementing its own broadcast or message forwarding mechanism.
The downside of this solution is that it makes the implementation of \textit{Messenger} more complex.\\
Although AMQP is generally used in client-server mode, implementations, such as RabbitMQ~\cite{rabbitmq}, propose a \textit{federation} mode
in which servers can be networked. In this mode, subscriptions are relayed to all the nodes in the federation and 
publications are routed to the right servers hosting subscribers. RabbitMQ takes care of all these operations.
This means that we do not finely master the exchanges between the different AMQP brokers, which can
be problematic in very constrained networks. 
Indeed, a message sent by a domain will be routed to every other domain interested. Eventually, the message will reach its 
destination but several copies will also arrive following different routes. The network footprint will thus increase, 
especially in a large interconnected network. 
Our current implementation of \textit{Messenger} uses the federation mode for simplicity reasons, but we plan to extend it with 
more efficient broadcast capabilities based, for instance, on a simple spanning tree.\\
\textit{Messenger} provides an open communication bus on top of which any agents can be plugged dynamically. It can
subscribe to topics published by other agents and start publishing on any topic. Note that security mechanisms
could be added to secure communication on a given set of topics or to filter publications from modules that have 
not been authorized.

\subsubsection{Agents}

To support QoS routing and reservation functionalities at inter-domain level, we have defined and implemented four 
main \textit{agents}. The \textit{Connectivity agent} is in charge of sharing with all the other domains the presence of peering links
with neighboring domains. This \textit{agent} works in an event-driven fashion as it sends information only if a new domain is discovered
or a peering link changes. This information is extracted and filled up from and into the \textit{Extended Database} of each controller, like
any other information received by \textit{agents}. This connectivity information will lately be specifically used by \textit{Path Computation} 
to locally take routing decisions. The \textit{Monitoring agent} periodically sends information on available bandwidth and latency between all the pairs of peering points to inform about the capability to support transit traffic in the domain. The \textit{Reachability agent} advertizes on an event basis the presence of hosts in domains so that they
become reachable. This service can be conceptually seen as an implementation of the Locator/Identifier Separation Protocol (LISP)~\cite{rfc6830} as it maintains at each controller a mapping between hosts and domains. The \textit{Reservation agent} takes care, like RSVP, of inter-domain flow setup, teardown and update request including application capability requirement such as QoS, bandwidth, latency, etc. 
All these requests are locally handled in each domain by the \textit{Service Manager}. 
\textit{Reservation agent} uses direct communications with neighbor domains along the paths that need to be created or maintained.\\
Each \textit{agent} publishes and consumes messages on a required subset of topics that they manage to ensure the consistency in the system. The exchanged information concern reachability (a list of reachable hosts in agent's domain), connectivity (a list of peering domains), and monitoring (the status of peering transit paths in terms of latency, bandwidth,...).
This way, each domain controller is able to build a view of the inter-domain network and have capabilities to perform routing, path reservation and manage SLAs.

\subsection{Interoperability issues}
 
The SDN domains managed by DISCO may communicate with other domains using classical IETF technologies.
To ensure this interoperability, like in the Google deployment B4, border nodes may have to send BGP messages to exchange
connectivity information. To manage this, an additional BGP agent should be added to the current architecture.\\
DISCO is agnostic from the SDN protocol and switches used. Despite the fact that our current implementation is OpenFlow-based, it could be integrated, for instance, with OnePK to manage CISCO equipment.

%% file: implem.tex
\section{DISCO Implementation}
\label{sec:implementation}

We have implemented DISCO on top of the open source OpenFlow controller Floodlight~\cite{floodlight}. 
The green hatched modules on Fig.~\ref{fig:architecture} have been directly taken from Floodlight's Java source code. We have developed in Java the other software modules, except the two SDN protocol drivers in yellow that are currently empty, to manage intra-domain and inter-domain.
Second, \textit{agents} located in the different domains use this control channel to receive and announce states of peering links, device locations, and path reservation requests.

\subsection{\textit{Messenger} implementation}

\textit{Messenger} is implemented like any other Floodlight application. 
It subscribes to receive Packet\_IN messages from the \textit{Core} module, can write its own Packet\_OUT messages,
calls and stores information in the extended database and reads the Floodlight configuration file at startup. 
It requires the following configuration parameters:
\textit{messaging\_server\_type}, \textit{messaging\_server\_listening\_port}, and \textit{agents\_list}.
The \textit{messaging\_server\_listening\_port} specifies the port where the \textit{Messenger} instance can be reached.
The \textit{messaging\_server\_type} determines which messaging driver to use, as our architecture allows using different AMQP implementations such as RabbitMQ or ActiveMQ. Our current implementation relies on a RabbitMQ 
driver using AMQP in federation mode. Additional optional parameters can also be specified.
\textit{Messenger} activates \textit{agents} from the list \textit{agents\_list}. \textit{Agents} are small classes that handle 
inter-domain exchanges to and from modules managing intra-domain flows. \\
\textit{Messenger} implements an extended version of LLDP (Link Layer Discovery Protocol), that we call Messenger-LLDP (M-LLDP), 
to discover neighboring domain controllers. M-LLDP messages are similar to regular LLDP messages but contain an option for OpenFlow.
An Organizationally Unique Identifier (OUI) has been allocated to OpenFlow by IEEE. 
 \textit{Messenger} sends these messages to announce its presence on border links where other OpenFlow domains may be reached,
 i.e., where switches that it manages have ports leading to unknown equipment. When a reply to a discovery message is received, 
 \textit{Messenger} establishes an AMQP connection with its peer and stops sending discovery messages on the border link.  
 Otherwise, \textit{Messenger} keeps sending periodically the following M-LLDP messages with information on how to reach him (a packet of 60 Bytes, which is a relatively low network footprint):

{\scriptsize
\begin{verbatim}
   0x7F (127 - LLDP's Custom TLV type)
      0x00 0x26 0xE1 (OpenFlow OUI) 
        0x17 (Messenger subtype)
          0x02 (controller ID)
          0x03 (switch ID) 
          0x04 (switch port) 
          0x05 (server IP)
          0x06 (server port)
          0x08 (server name)
\end{verbatim}}

\textit{Messenger} offers a publish/subscribe communication channel for inter-domain exchanges between \textit{agents}.
\textit{Messenger} uses two special topics for its basic operations. 
First, a topic named \textit{ID.*.*} is created, ID being the identifier of the controller. This topic allows other controllers 
to directly send messages to it. This is used, for instance, for bandwidth reservation requests. 
Second, a topic named \textit{general.*.*} enables to communicate with all the other controllers in the federation.
For example, it is used when a controller wants to leave a federation. This deletes the logical link between itself and the other controller 
and warns the \textit{agents} to stop sending messages to this particular controller. \\
\textit{Messenger} uses drivers to communicate with the implementation of AMQP running on the machine. Each AMQP driver must support the following set of functions:
\begin{enumerate}
\item subscribe (topic) and unsubscribe\_topic (topic): add and delete a \textit{topic} from the topic list that the node is interested to receive.
\item pair (neighbor controller ID) and unpair (neighbor controller ID): create and delete a inter-domain control channel with a \textit{neighbor controller}.  
This function tunes the subscriptions so that a node receives or not information from this \textit{neighbor controller}.
\item send (topic, message): send a \textit{message} on a specific \textit{topic} to the federation.  
\end{enumerate}
\textit{Messenger} also uses Keep-Alive messages every 500ms to test the presence of neighboring controllers. 
In case of a controller failure, the absence of 3 successful contiguous Keep-Alive responses will trigger a procedure to mitigate this failure. \\
The \textit{Messenger} application has been conceived to be easily extensible. Without altering the \textit{core} classes of Floodlight, 
a developer can improve it either by providing a driver for a different implementation of AMQP extending the abstract 
\textit{MessengerDriver} class or add functionalities by adding another \textit{agent}.
The topic format that we use is also highly flexible as every \textit{agent} can define its own topic. Furthermore, wild-cards can be used to make subscriptions lighter and better manageable by developers.

\begin{figure}[!t]
\centering
\includegraphics[width=0.9\linewidth]{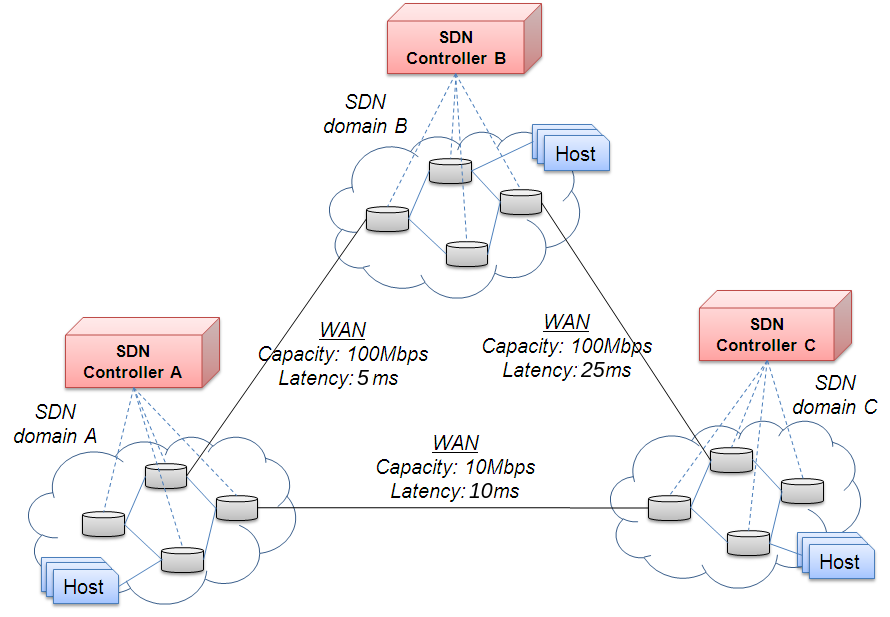}
\caption{Multi-domain SDN Topology.}
\label{fig:topology}
\end{figure} 

\subsection{Agents implementation}

\textit{Agents} use \textit{Messenger} to exchange information with neighboring domains.
We have implemented four agents: \textit{Monitoring}, \textit{Reachability}, \textit{Connectivity} and \textit{Reservation} (see Sec.~\ref{sec:inter}). 
They all publish on specific topics such as \textit{monitoring.ID.bandwidth.2s} that the monitoring \textit{agent} located at the controller 
with identifier ID advertizes every other second the remaining bandwidth that it can offer to transit traffic.\\
Upon reception of information from \textit{agents} in neighboring domains, the local \textit{agents} store them in the extended 
Floodlight database. This information is then used by local modules to take decisions on
flows. These decisions are generally the outgoing peering link to choose for a given flow.\\
The \textit{Reservation agents} implements a RSVP-like reservation protocol to provision end-to-end resources.
\textit{Agents} thus exchange reservation requests and responses with flow descriptors. Messages can be directly sent to the next domain controller on a path with the \textit{ID.*.*.} topic.\\
\textit{Messenger} and its dependencies (\textit{agents}, drivers, \ldots) were written with just over 2400 lines of Java code. There was minimal intervention in the legacy code, except in the GUI and the Extended Database to represent remote domains. The intra-domain modules written beforehand to extend Floodlight amount to almost 12,000 lines of code. While running, \textit{Messenger} only adds around 14MB to the total memory used by Floodlight, which is about 100MB\footnote{This value depends mainly on the heap size limit allocated to the JVM. It can be as low as 64MB up to several hundred Megabytes.}.

%% file: evaluation.tex
\section{Evaluation}
\label{sec:evaluation}

In this section, we present how we assessed the capabilities of DISCO. 
In addition to classic functions such as QoS routing, reservation and pre-emption, DISCO 
aims at being resilient to disruptions both on the control plane (e.g., controller failure, inter-controller communication failure) 
and on the data plane (e.g., inter-domain link failure). We have thus define three use cases that enable the evaluation of these features.

\begin{figure}[!t]
        \centering
		\includegraphics[width=0.8\linewidth]{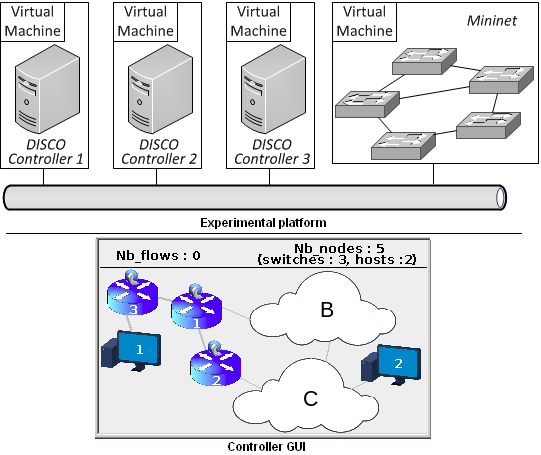}
        \caption{Experimental setup (top) and controller A's GUI (bottom).}
        \label{fig:setup}
\end{figure}

\subsection{Testbed and setup}

Fig.~\ref{fig:topology} presents the network topology considered in the performance evaluation. 
Each network domain A, B and C is managed by a local DISCO controller, which coordinates with its neighbor DISCO controllers. 
This setup is representative from a typical enterprise network where several sites (edge networks or datacenters) are interconnected with different WANs. The hosts connected to the network domains can be either user terminals or virtual machines (VM).\\
The testbed is enclosed in a private cloud as shown in Fig.~\ref{fig:setup} (top). 
The network is emulated using \textit{Mininet}~\cite{mininet}, 
a tool used to create rich topologies and instantiate Open vSwitch~\cite{openvswitch} switches and virtual hosts. \textit{Mininet} is 
hosted on a dedicated VM and the controllers are hosted on separate VM. The different link latencies and bandwidths 
are enforced using Linux's \textit{tc} command. This setup allows us a fine control on the network. 
Fig.~\ref{fig:setup} (bottom) also presents the graphical user interface of controller A showing that it has the knowledge of all the
switches it manages, of all the other domains, namely B and C, and of the different hosts, local and remote.\\

\subsection{Use Case 1: Adaptive information exchange}

\begin{figure}[!b]
        \centering
			\includegraphics[width=0.9\linewidth]{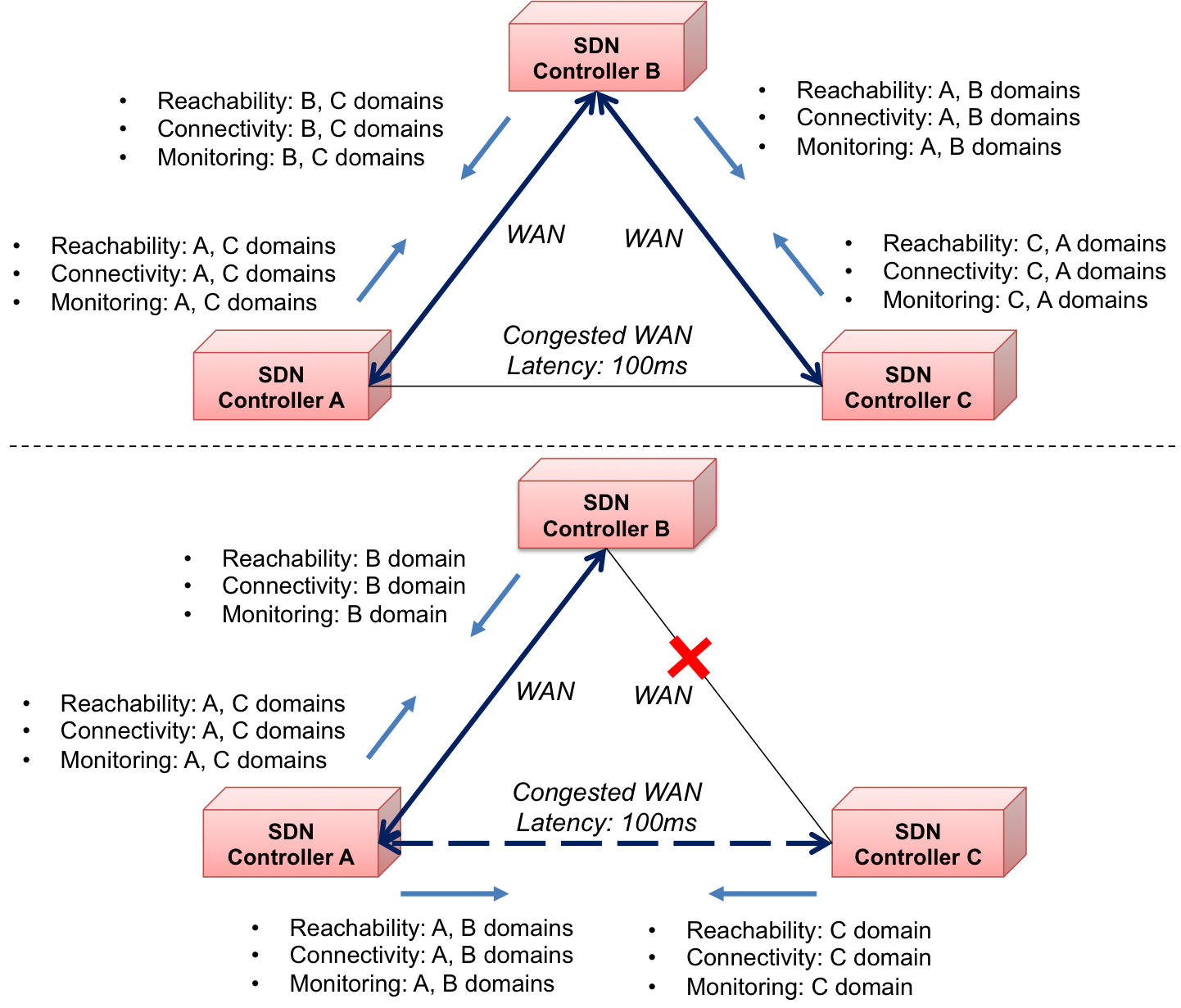}
        \caption{Adaptable information exchange: (top) Congested situation: DISCO controllers use high capacity inter-domain links to exchange information, (bottom) Inter-domain link disruption: DISCO controllers adapt the content and the frequency of their information exchanges.}
        \label{fig:monitoring}
\end{figure}
\begin{figure*}[!th]\centering
\subfigure[A $\rightarrow$ B]{\label{traffic.a.b}\includegraphics[width=5.9cm]{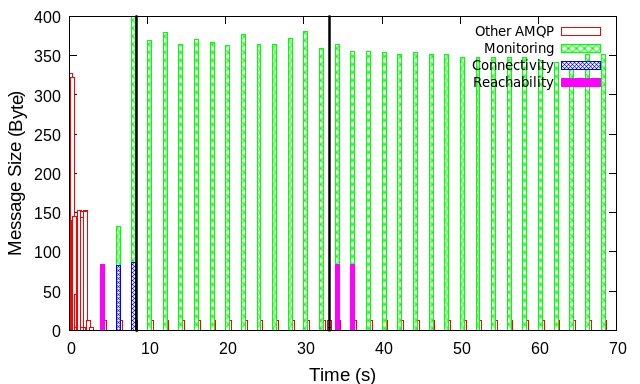}}
\subfigure[A $\rightarrow$ C]{\label{traffic.a.c}\includegraphics[width=5.9cm]{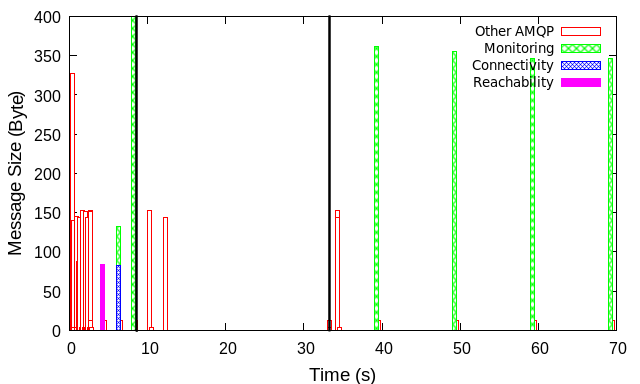}}
\subfigure[B $\rightarrow$ C]{\label{traffic.b.c}\includegraphics[width=5.9cm]{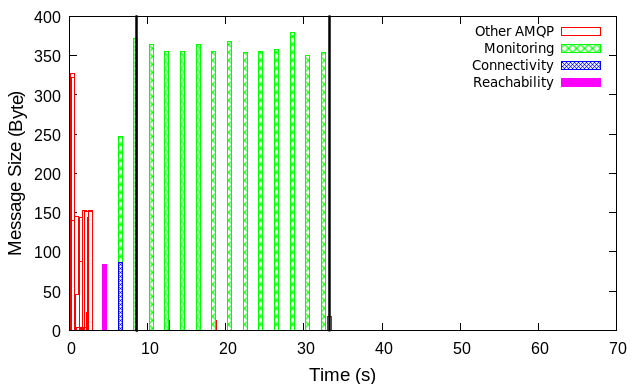}}
\subfigure[B $\rightarrow$ A]{\label{traffic.b.a}\includegraphics[width=5.9cm]{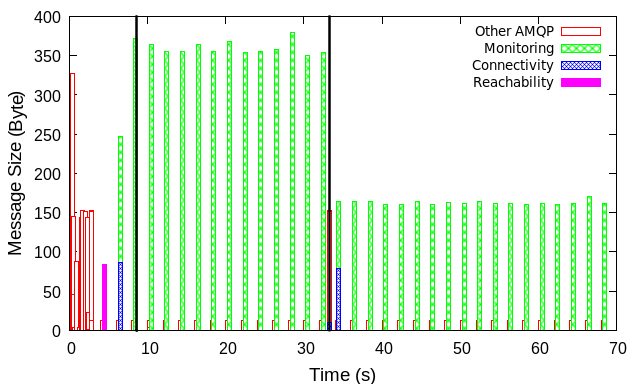}}
\subfigure[C $\rightarrow$ A]{\label{traffic.c.a}\includegraphics[width=5.9cm]{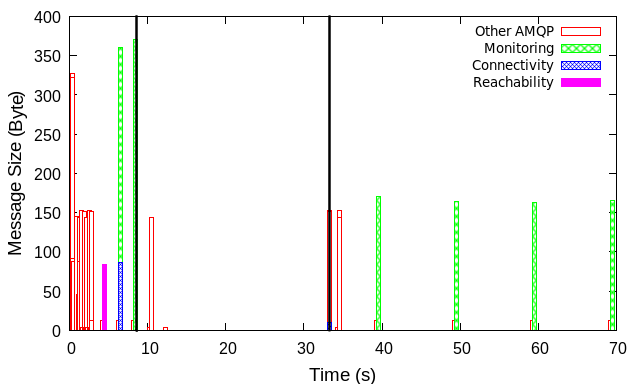}}
\subfigure[C $\rightarrow$ B]{\label{traffic.c.b}\includegraphics[width=5.9cm]{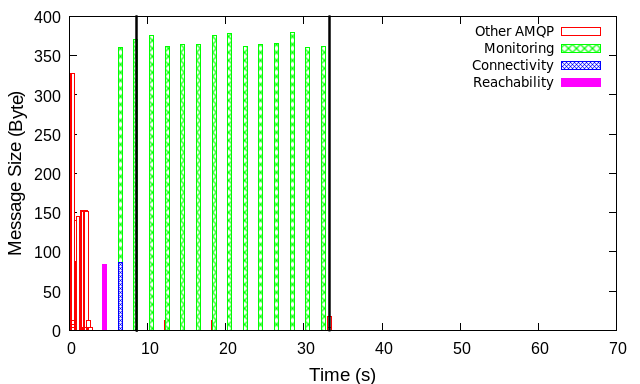}}
\caption{\label{fig:topology-monitoring1.results} Adaptive monitoring information exchange traffic on the different links. Packets come from the different agents and AMQP itself. At $t=9s$, the bootstrap and discovery phases end. At $t=33s$, the link $B \leftrightarrow C$ is cut off.}
\vspace{-3mm}
\end{figure*}
In this scenario, we show how the exchanges in the control plane can self-adapt to the network conditions. 
In order to reduce the network footprint of control information exchanged between domains, \textit{agents} adopt a twofold strategy: 
(1) they identify alternative routes to offload traffic from weak outgoing interconnections (e.g., low-bandwidth satellite connection, congested link), and (2) they reduce the frequency of control messages for these links if they do not find an alternative route. Each \textit{Monitoring agent} usually sends information every $2s$. This period increases to $10s$ for weak interconnections. 
The \textit{Connectivity} and \textit{Reachability agents} also send their information using alternative routes whenever possible. 
However, contrary to the \textit{Monitoring agents}, they only exchange messages in a reactive manner, that is when an event occurs.\\
Upon bootstrap and discovery, the three controllers reach the situation described on top of Fig.~\ref{fig:monitoring}. 
In this scenario, the link between the domains A and C is congested. Its latency equals $\ge 50ms$. B is thus relaying control information for A and C in order to offload the congested link. In case the inter-domain link between B and C fails, \textit{Monitoring agents} reconfigure themselves to the situation presented at the bottom of Fig.~\ref{fig:monitoring} where monitoring traffic is passed through the weak link $A \rightarrow C$, but at a lower frequency.\\
Fig.~\ref{fig:topology-monitoring1.results} presents the evaluation we have conducted to show how the DISCO control plane adapts to the  
nominal situation and when the inter-domain link between B and C fails. This figure presents
the link utilization in both directions right after the controllers discover
each other and start exchanging AMQP messages. Each bar represents the TCP payload size of received packets per 
category\footnote{In the current implementation, controllers exchange JSON messages for ease of development and integration. However,   
compression is planned in future releases.}.
This experimental scenario can be split up into three phases: 
\begin{enumerate}
\item \textit{Network discovery} till $t=9s$ where controllers exchange their knowledge about hosts and the inter-domain network topology. AMQP is particularly active during this phase because the brokers have to create the federations and subscribe to the different topics.
In this phase the monitoring has already started but is not yet adapted to weak links.
\item \textit{Monitoring adaptation} from $t=9s$ to $t=33s$ where agents have discovered a weak link and adapt their behavior accordingly. 
We observe on Fig.~\ref{traffic.a.c} and Fig.~\ref{traffic.c.a} that monitoring is shot down after $t=10s$ as the link $C \leftrightarrow A$ is weak (congested), 
while monitoring traffic increases on Fig.~\ref{traffic.b.c} and Fig.~\ref{traffic.b.a}. 
\item \textit{Failure recovery} starting right after we cut the link between B and C at $t=33s$. Information is transmitted over the link 
between A and C, but with an adapted frequency as shown in Fig.~\ref{traffic.c.a} and Fig.~\ref{traffic.a.c}. Monitoring traffic sent over $B \rightarrow A$ decreases as information about $B \leftrightarrow C$ is no longer necessary. 
\end{enumerate}
We additionally tested what would happen if a controller fails entirely. \textit{Messenger} has a built-in feature whereupon if a controller fails `gracefully', it can warn its neighbors so that they can prepare for the failure.
Otherwise, the Keep-alive system will warn a controller if its neighbor is no longer reachable. In that case, the logical control plane links are severed, no messages are carried any more toward this controller and other failure mitigation processes occur (e.g., if the fallen domain was used to exchange messages between two domain, they will reconnect by other path if available).

\subsection{Use Case 2: Resource reservation and pre-emption}

\begin{figure}[!b]
\centering
\includegraphics[width=0.90\linewidth]{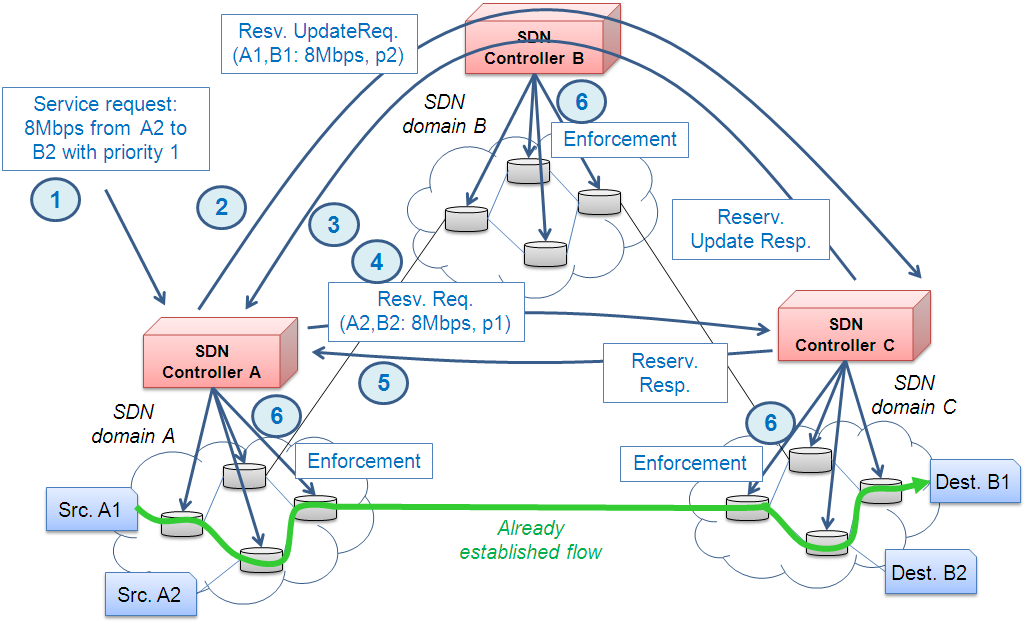}
\caption{Pre-emption case where a service request on node A triggers a resource reservation on link $A \leftrightarrow C$ 
where a lower priority flow is already established.}
\label{fig:preemption}
\end{figure}

\begin{figure}[!b]
\centering
\includegraphics[width=0.9\linewidth]{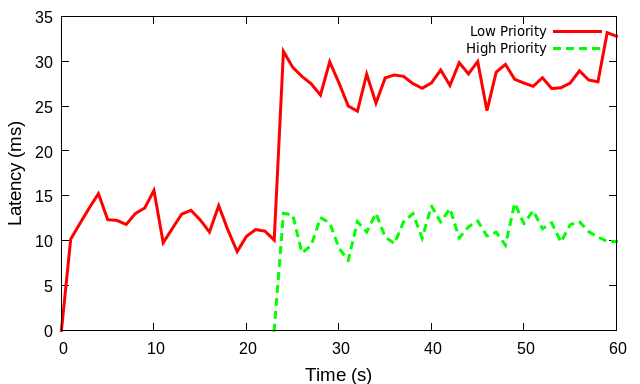}
\caption{Impact on flow latency when the link $A \leftrightarrow C$ is pre-empted by a higher priority flow.}
\label{fig:reservation2}
\end{figure}

This scenario shows how DISCO manages resource reservation thanks to the \textit{Service Manager} and the \textit{Reservation agent}. An ongoing flow from host A1 to host C1 using 8 Mbits/s has already been set up on the link between A and C (Fig.~\ref{fig:preemption}). The routing decision was made by the path computation module of A considering the available inter-domain capacity. At $t=25s$ a new flow with a higher priority and a 8 Mbits/s bandwidth constraint from host A2 to host C2 initiates a reservation with a strict latency constraint of 15 ms maximum. This reservation request is handled on the fly when the first packet arrives. However, it could have been requested through the northbound controller API in advance. Fig.~\ref{fig:preemption} depicts the flow of control messages involved in this reservation process. The \textit{Reservation agent}, using the \textit{Path Computation} module finds out that the only suitable inter-domain path for this priority request goes through the direct link between A and C, that is the one already used by the non-priority flow from A1 to C1. The agent thus initiates a reservation to deviate the non-priority flow through domain B, which is a longer path. Then, the resources on the direct link between domain A and domain C having been released by this operation, the agent triggers a reservation for the priority flow.\\
Fig.~\ref{fig:reservation2} presents the latency for both flows. The reservation procedure took less than $500ms$. We can observe that a transitory impact occurs on the latency of the non-critical flow. When the reservation is agreed and the low-priority flow rerouted, the latency of this latter flow is significantly increased as a longer route is used.

\subsection{Use Case 3: Virtual Machine migration}

\begin{figure}[!h]
\centering
\includegraphics[width=0.9\linewidth]{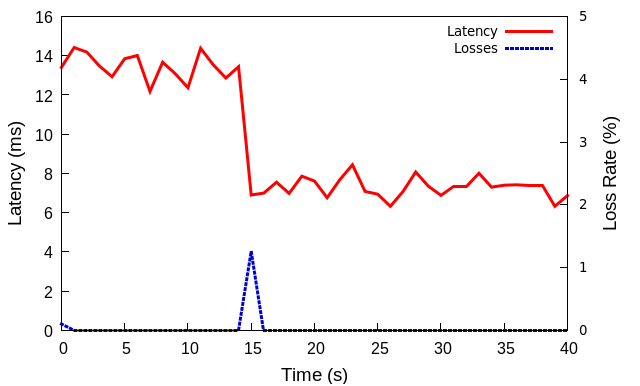}
\caption{Impact on flow latency and loss rate when the destination host $2$ is moved from domain $C$ to $B$.}
\label{fig:migration3}
\vspace{-2mm}
\end{figure}

In this last scenario, a virtual machine is migrated from one domain to another by a cloud management system. 
Upon detection at layer 3 that a new IP appears in a domain, the \textit{Reachability agent} sends an update message to 
announce that it now manages this new host. Other controllers update the mapping that they maintain locally between hosts and 
domains so that their path computation module can route flows towards this new location. To speed up the convergence of the handover,
they also immediately update all the rules related to this host in the switches that they manage if a flow going toward this IP was already established.
To test this case, we launched a 10 Mbits/s UDP flow from a host located in domain A to another situated in C during 40s. 
According to the values presented in Fig.~\ref{fig:topology}, this flow is taking all of the available bandwidth on the 
$A \leftrightarrow C$ link whose lantecy is equal to 10ms. If the cloud management system moves the Virtual Machine from domain 
C to B, C will announce that it is no longer capable of reaching the VM while B reports that it is now handling it following 
the migration. These messages, along with the reconfiguration of the mapping between host and domain, will allow A 
to dynamically reroute the flow to its new destination.\\
Fig.~\ref{fig:migration3} shows the performance in terms of service interruption and network metrics. The VM is now 
reachable through the $A \leftrightarrow B$ link which has a lower latency. We can see the latency drops at the 
moment of the change in the flow's path. We ran this experiment ten times to get average values. The switch is not 
instantaneous; it takes on average $117 ms$ with some packet losses ($1.26\%$ in average).
In any case, this experiment shows the adaptability of DISCO in a environment where end-hosts can move between domain and their communication can be seamlessly rerouted by the controllers.

%% file: conclusion.tex
\section{Conclusions and perspectives}
\label{sec:conclusion}

In this paper, we have proposed DISCO, an open \emph{DIstributed SDN COntrol plane} for multi-domain networks. 
It relies on a per domain organization where each DISCO controller is in charge of an 
SDN domain and provides a unique lightweight and highly manageable control channel used by \textit{agents} that can be dynamically plugged into the different domain controllers. The agents that we have developed share between the domains aggregated network-wide information and hence provide end-to-end network services. We demonstrated how DISCO dynamically adapts to heterogeneous network topologies while being resilient enough to survive to disruptions and attacks and 
providing classic functionalities such as end-point migration and network-wide traffic engineering.
Contrary to state of the art distributed SDN control planes, DISCO well discriminates heterogeneous inter-domain links such as 
high-capacity MPLS tunnels and SATCOM interconnections. We have implemented DISCO on top of the 
Floodlight OpenFlow controller ~\cite{floodlight} and the AMQP protocol~\cite{amqp}. We have evaluated its functionalities according to three use cases: 
inter-domain topology disruption, end-to-end priority service request and virtual machine migration.\\
Future work along these lines include the organization of the controllers. In our current implementation,
all the controllers have equal role and can communicate with neighboring ones. In case of high degree
network topologies, one could choose to silent some of the inter-controller links while keeping 
good performances. Also of interests, controllers could dynamically regroup in coherent 
clusters so that one of them can take decisions in a centralized manner. This way, 
suboptimal allocations such as the one found by a distributed RSVP-like protocol would be avoided.
Finally, we would like to enrich DISCO with additional resilient and recovery mechanisms so that a controller can on the fly take the control of switches from a neighbor domain in case of failure.